# Anisotropy in the electronic structure of V$_2$GeC investigated by soft x-ray emission spectroscopy and first-principles theory

Martin Magnuson[1,5], Ola Wilhelmsson[2], Maurizio Mattesini[3], Sa Li[1,4], Rajeev Ahuja[1], Olle Eriksson[1], Hans Högberg[5], Lars Hultman[5] and Ulf Jansson[2]

[1]*Department of Physics, Uppsala University, P. O. Box 530, S-751 21 Uppsala, Sweden.*

[2]*Department of Materials Chemistry, The Ångström Laboratory, Uppsala University, P.O. Box 538 SE-75121 Uppsala.*

[3]*Departamento de Física de la Tierra, Astronomía y Astrofísica I, Universidad Complutense de Madrid, E-28040, Spain.*

[4]*Department of Physics, Virginia Commonwealth University, Richmond, Virginia 23284-2000, USA.*

[5]*Department of Physics, IFM, Thin Film Physics Division, Linköping University, SE-58183 Linköping, Sweden.*

## Abstract

The anisotropy of the electronic structure of ternary nanolaminate V$_2$GeC is investigated by bulk-sensitive soft x-ray emission spectroscopy. The measured polarization-dependent emission spectra of V $L_{2,3}$, C $K$, Ge $M_1$ and Ge $M_{2,3}$ in V$_2$GeC are compared to those from monocarbide VC and pure Ge. The experimental emission spectra are interpreted with calculated spectra using *ab initio* density-functional theory including dipole transition matrix elements. Different types of covalent chemical bond regions are revealed; V 3$d$ - C 2$p$ bonding at -3.8 eV, Ge 4$p$ - C 2$p$ bonding at -6 eV and Ge 4$p$ - C 2$s$ interaction mediated via the V 3$d$ orbitals at -11 eV below the Fermi level. We find that the anisotropic effects are high for the 4$p$ valence states and the shallow 3$d$ core levels of Ge, while relatively small anisotropy is detected for the V 3$d$ states. The macroscopic properties of the V$_2$GeC nanolaminate result from the chemical bonds with the anisotropic pattern as shown in this work.

# 1 Introduction

In the last years, a group of ternary nanolaminated carbides and nitrides denoted M$_{n+1}$AX$_n$ (MAX phases), where n=1, 2, and 3 refers to 211, 312, and 413 crystal structures, respectively, has attracted much attention [1]. Here, M denotes an early transition metal, A is a p-element, in groups IIIA and IVA, and X is either carbon and nitrogen [2]. The MAX-phase compounds exhibit a unique combination of metallic and ceramic properties, such as low density, high strength and stiffness at high temperatures, resistance to oxidation and thermal shock, in addition to high electrical and thermal conductivity [1]. The hexagonal crystal structures of the ternary MAX-phase compounds (space group P63/mmc) are closely related to the binary monocarbides and nitrides (MX). In spite of the structural similarities with the binaries, MAX-phases exhibit very different chemical and physical properties. This is directly attributed to the anisotropic nanolaminated crystal structure with alternating strong M-X bonds and much weaker M-A bonds [3, 4]. The binary $M_{n+1}X_n$ nanoscale blocks are interleaved with square-planar monolayers of pure A-elements, where the X-atoms (C or N) fill the octahedral sites between the M-atoms. The A-elements are located at the center of trigonal prisms that are larger than the octahedral X sites [5, 6, 7, 8].

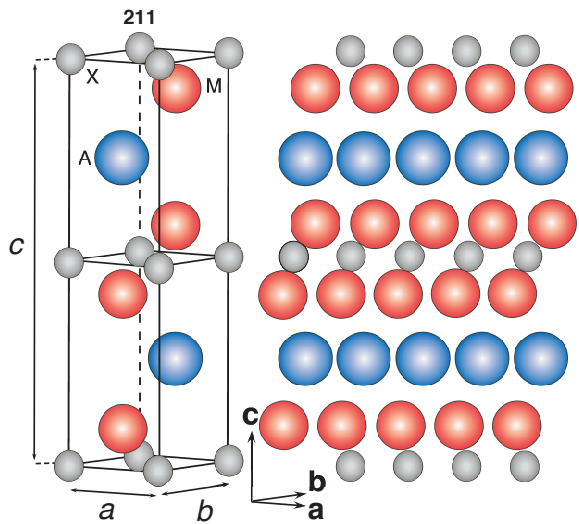

**Figure 1:** The hexagonal crystal structure of the 211 phase ($V_2GeC$). There is one A (Ge) layer for every second layer of M (V) in $V_2GeC$. The M (V) atoms have chemical bonds to both X (C) and A (Ge). The lengths of the measured (calculated) $a$ and $c$-axis of the hexagonal unit cell of $V_2GeC$ are 2.99 (3.01) Å and 12.28 (12.18) Å, respectively. The length of the $a$-axis in cubic VC is 4.14 (4.16) Å.

Macroscopic properties such as electrical and thermal conductivity and elasticity of ternary carbides and nitrides are known to strongly depend on the valence-electron concentration of the M-element. Information about the internal electronic structure and the anisotropic pattern of the chemical bonds is therefore invaluable but generally difficult to obtain experimentally. Previous experimental investigations of the electronic structure and chemical bonding of the 211-crystal structure mainly include Ti-based compounds such as $Ti_2AlC$ [9] and $Ti_2AlN$ [10]. For investigating the trend in the anisotropy of the chemical bond scheme, the nature of the electric and thermal conduction and elastic properties, it is of great scientific interest to replace Ti with a heavier 3$d$ element such as V. For this purpose, we have synthesized and investigated epitaxial $V_2GeC$ and $VC_x$ films grown on $Al_2O_3(000l)$ single crystals. Figure 1 shows the crystal structure of $V_2GeC$ with thermodynamically stable binary V-C-V blocks separated by softer V-Ge-V blocks with weaker bonds. Intercalation of Ge monolayers into the VC matrix implies that the strong V-C bonds are broken up and replaced by the weaker V-Ge bonds. Thus, in $V_2GeC$, every second layer of C atoms in VC has been replaced by a Ge monolayer, in effect resulting in understoichiometric VC. The VC blocks are then twinned with the Ge layers acting as mirror planes. The 211 crystal structure of $V_2GeC$ contains anisotropic V atoms (octahedral coordination in-plane and trigonal prismatic coordination out-of-plane) with chemical bonds both to C and Ge atoms while stoichiometric VC monocarbide contains isotropic V atoms (octahedral coordination only) which only bond to C.

In the present paper, we investigate the anisotropy in the nanolaminated internal electronic structure of $V_2GeC$ in comparison to monocarbide VC and pure Ge. Bulk-sensitive and element-specific soft x-ray emission (SXE) spectroscopy with selective excitation energies around the V 2$p$, C 1$s$, Ge 3$s$ and Ge 3$p$ absorption thresholds is utilized. The involvement of both valence levels and different kind of core levels, makes it possible to study each of the constituent elements separately in the ternary compound. Changing the polarization of the electric field vector (**E**), of the incoming radiation, effectively enables obtaining directional information about the chemical bonding which cannot easily be obtained with other methods. We present the first electronic structure investigation of $V_2GeC$ which is particularly important for understanding and predicting the macroscopic properties. The SXE spectra are interpreted in terms of pDOS weighted by the dipole transition matrix elements.

# 2 Experimental

## 2.1 Deposition of the $V_2GeC$ and VC films

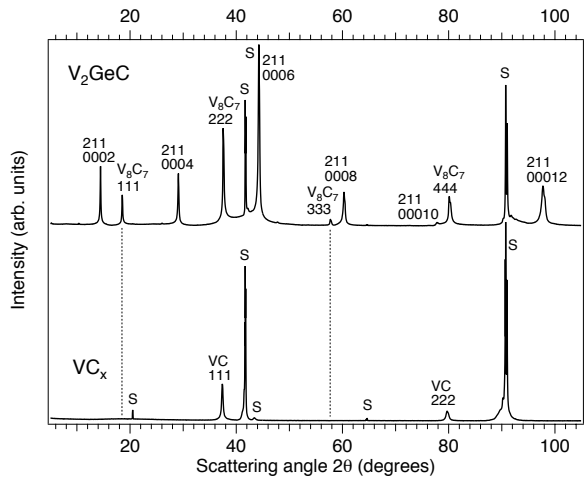

**Figure 2:** X-ray diffractograms (XRD) from the $V_2GeC$ (0001) and VC(111) thin film samples grown at $700^o$ C and $500^o$ C, respectively. The diffraction peaks at $2\Theta = 18.4^o$ and $57.4^o$ indicated by the vertical dashed lines are due to a minor fraction of coherent $VC_x$ inclusions.

Figure 2 (top) shows $\Theta$-$2\Theta$ x-ray diffractograms (XRD) of the deposited $V_2GeC$ and $VC_x$ films. The $V_2GeC$ (000$l$) (5000 Å thick) and $VC_x$(111) ($x \sim$ 0.875, 2000 Å thick) films were epitaxially grown on $Al_2O_3$ (000$l$) substrates at $700^o$ C and $500^o$ C, respectively, by dc magnetron sputtering. Elemental targets of V, C and Ge, and a 3.0 mTorr Ar discharge were used [11]. Comparing the diffractograms of the $V_2GeC$ (Fig. 2, top) and $VC_x$ (bottom) films, peaks of {000$l$} type from the $V_2GeC$ MAX-phase are observed together with $VC_x$($lll$) and substrate peaks (S). The $VC_x$ phase has a cubic NaCl (B1-type) crystal structure and is known not to exhibit any diffraction peaks at $2\Theta = 18.4^o$ and $57.4^o$ as shown in the $VC_x$ diffractogram at the bottom of Fig. 2. However, at substoichiometric compositions and elevated temperatures, a disorder-order transformation occurs. The ordered structure consists of 8 B1 type unit cells and has a total superstructure composition of $V_8C_7$ (i.e. $VC_{0.875}$) [12]. In the $V_8C_7$ superstructure, the diffraction peaks at $2\Theta = 18.4^o$ and $57.4^o$ are allowed as observed in the diffractogram at the top of Fig. 2. The relative intensity of the $V_8C_7$ superstructure peaks compared to other $VC_x$ peaks increase with increasing deposition temperature [11]. This suggests that the $V_2GeC$ MAX-phase film contains a small mixture of disordered $VC_x$ and ordered $V_8C_7$ inclusions. In the following we will simply refer to the minor fraction of binary $VC_x$ monocarbide inclusions as $VC_x$. The formation of the $VC_x$ inclusions is due to the very narrow homogeneity range and the growth mechanism during the sputtering process. Minor fluctuations in the surface composition during the crystal growth lead to more favorable formation of $VC_x$ inclusions, segregation and enrichment of Ge at the surface, causing renucleation. These self-regulated growth fluctuations have also been observed in other MAX-phase systems such as Ti-Al-C [13]. In this case, it is not possible to synthesize films totally free from coherent $VC_x$ inclusions.

For the $V_2GeC$ film, x-ray pole figures show that the $VC_x$ inclusions are coherent with the MAX-phase showing an out-of-plane orientation $VC_x(lll)//V_2GeC(000l)$. The fact that the $V_2GeC$ diffractogram (top) shows mainly $V_2GeC$ of $\{000l\}$-type suggests a highly textured or epitaxial MAX-phase film. X-ray pole figures verified that the growth indeed was epitaxial with an in-plane orientation of $V_2GeC[10\bar{1}0]//Al_2O_3[21\bar{3}0]$ and an out-of-plane orientation of $V_2GeC(000l)//Al_2O_3(000l)$. The values of the a-axis and c-axis were determined to 2.99 and 12.28 Å, respectively, by reciprocal space mapping. X-ray photoelectron spectroscopy (XPS) analyses of the $V_2GeC$ and $VC_x$ films using a PHI Quantum instrument showed after 60 s of Ar-sputtering a constant composition. Except for a thin surface oxidation of approximately 5nm, no contamination species such as oxygen was detected in the bulk of the films by XPS depth profiling.

## 2.2 X-ray emission and absorption measurements

The soft x-ray absorption (SXA) and SXE measurements were performed at the undulator beamline I511-3 at MAX II (MAX-lab National Laboratory, Lund University, Sweden), comprising a 49-pole undulator and a modified SX-700 plane grating monochromator [14]. The SXE spectra were measured with a high-resolution Rowland-mount grazing-incidence grating spectrometer [15] with a two-dimensional multichannel detector with a resistive anode readout. The V $L$ and C $K$ SXE spectra were recorded using a spherical grating with 1200 lines/mm of 5 m radius in the first order of diffraction. The Ge $M_1$ and $M_{2,3}$ SXE spectra were recorded using a grating with 300 lines/mm, of 3 m radius in the first order of diffraction. The SXA spectra at the V $2p$ and C $1s$ edges were measured with 0.12 eV and 0.06 eV resolutions using total electron yield (TEY). In SXA, the selection rules enables the excitation cross section into a particular unoccupied orbital to be maximized if the incident **E**-vector is oriented parallel to the orbital axis, and minimized if it is perpendicular [16].

During the V $L_{2,3}$, C $K$, Ge $M_1$ and $M_{2,3}$ SXE measurements, the resolutions of the beamline monochromator were 0.45, 0.2, 0.2, and 0.02 eV, respectively. The SXE spectra were recorded with spectrometer resolutions of 0.42, 0.2, 0.2, and 0.02 eV, respectively. All measurements were performed with a base pressure lower than $5\times10^{-9}$ Torr. In order to minimize self-absorption effects [17], the angle of incidence was 20$o$ from the surface plane during the SXE measurements. The SXE spectra were recorded in two different geometries: with the polarization vector of the incoming x-rays either parallel to the c-axis and the scattering plane (**E**|**c**, *p*-polarization) or parallel to the surface plane (**E**|**a,b**, *s*-polarization), see Fig. 1. In both cases, the detector was placed in a direction perpendicular to the incident x-ray beam, either in the plane of the synchrotron orbit or perpendicular to it. The first geometry (horizontal position of the detector and **E**|**c**) has the advantage that elastic scattering is supressed since Rayleigh scattering does not occur in the direction of the electrical field vector of the incident radiation [16].

# 3 Computational details

## 3.1 Calculation of the soft x-ray emission spectra

The SXE spectra were calculated within the single-particle transition model by using the augmented plane wave plus local orbitals (APW+lo) band structure method [18]. Exchange and correlation effects were described by means of the generalized gradient approximation (GGA) as parameterized by Perdew, Burke and Ernzerhof [19]. We previously successfully modelled SXE for MAX-phases using similar methodology, and obtained excellent agreement with experiment [20, 21, 9, 10], giving credence to the present predictions for the $V_2GeC$ system.

The total energy was first converged against the k-point integration (21×21×4) whereafter the theoretical emission spectra within the electric-dipole approximation were computed. We utilized a plane wave cut-off corresponding to $R_{MT}*K_{max}$=8.0 and a muffin-tin radius of 1.90 atomic units (a.u.) for V, Ge, and C. The core-hole lifetimes employed in these calculations were 0.40 eV, 0.12 eV, 2.0, and 2.0 eV for the V $2p$, C $1s$, Ge $3s$, and $3p$ edges, respectively. Finally, we achieved a direct comparison of the calculated spectra with the measured data by including an instrumental broadening in the form of Gaussian functions (see section IIB). The final state lifetime broadening was accounted for by a convolution with an energy-dependent Lorentzian function with a broadening increasing linearly with the distance from the Fermi level according to the function $a+b(E-E_F)$, where the constants $a$ and $b$ were set to 0.01 eV and 0.05 (dimensionless) [22].

## 3.2 Balanced crystal orbital overlap population (BCOOP)

In order to study the chemical bonding of the $V_2GeC$ compound, we calculated the BCOOP function by using the full potential linear muffin-tin orbital (FPLMTO) method [23]. In these calculations, the muffin-tin radii were kept as large as possible without overlapping each other (V=2.2 a.u, Ge=2.2 a.u, and C=1.6 a.u.). To ensure a well-converged basis set, a double basis with a total of four different $\kappa^2$ values were used. For V, we included the $4s$, $4p$ and $3d$ as valence states. To reduce the core leakage at the sphere boundary, we also treated the $3s$ and $3p$ core states as semi-core states. For Ge, $4s$, $4p$ and $4d$ were taken as valence states. The resulting basis formed a single, fully hybridizing basis set. This approach has previously proven to give a well-converged basis [24]. For the sampling of the irreducible wedge of the Brillouin zone, we used a special-k-point method [25] and the number of k points were 1000 for $V_2GeC$ and 1728 for VC in the self-consistent total energy calculation. In order to speed up the convergence, a Gaussian broadening of 20 mRy width was associated with each calculated eigenvalue.

# 4 Results

## 4.1 V $L_{2,3}$ x-ray emission

Figure 3 (top, right curves) shows V $2p_{3/2,1/2}$ SXA spectra of $V_2GeC$ and VC measured to identify the photon energies of the absorption intensity peak maxima. The SXA measurements mainly probe the unoccupied V $3d_{xy}$, $3d_{x^2-y^2}$ in-plane orbitals. In the middle of Fig. 3, V $L_{2,3}$ SXE spectra of $V_2GeC$ and VC are shown. The **E|a,b** excited SXE spectra predominantly probe the occupied V $3d_{xy}$, $3d_{x^2-y^2}$ in-plane orbitals. With **E|c** polarization, there is more influence from

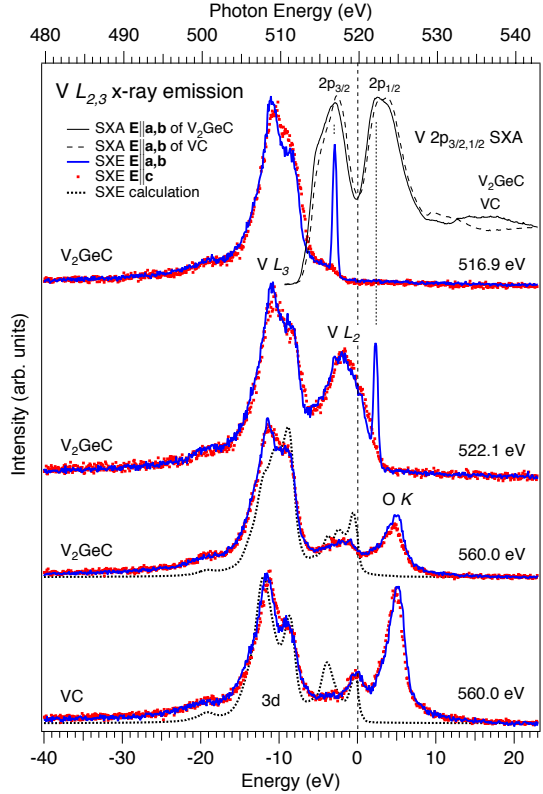

**Figure 3:** The top-right curves show V $2p_{3/2,1/2}$ SXA spectra measured at normal incidence (**E|a,b**), normalized to each other below and far above the absorption step edge. In the middle (top to bottom), a series of V $L_{2,3}$ SXE spectra of $V_2GeC$ and VC are shown. The SXE spectra are excited with the polarization vector both parallel (**E|a,b**) and perpendicular (**E|c**) to the surface plane at 516.9, 522.1 (resonant) and 560.0 eV (nonresonant). For each excitation energy, the spectra were normalized to the time and incoming photon flux by the measured current from a gold mesh in the photon beam. The excitation energies of the resonant SXE spectra with **E|a,b** polarization are indicated by the strong elastic peaks determining the total resolution (FWHM=0.6 eV). The spectra are plotted both on a photon energy scale (top) and a common energy scale relative to the $E_F$ (bottom) using the $2p_{1/2}$ ($2p_{3/2}$) core-level XPS binding energy of 519,9 eV (512.3 eV) for $V_2GeC$. For the nonresonant spectra at the bottom, calculated isotropic SXE spectra are shown by the dotted curves.

the out-of-plane V $3d_{z^2}$ orbitals. In the V $L_{2,3}$ spectra of $V_2GeC$ excited at 560.0 eV, four peaks are observed at -2 eV, -9 eV, -11 eV and -19 eV below $E_F$. For the non-resonant spectra (560 eV) of both $V_2GeC$ and VC, a O $K$ emission peak at 525 eV on the photon energy scale at the top is observed. This peak appears for all excitation energies above the O $1s$ absorption edge at 543.1 eV and originates from a thin surface oxide layer on the samples. As the excitation energy is tuned resonantly to the V $2p_{3/2}$ and $2p_{1/2}$ absorption maxima at 516.9 and 522.1 eV, the V $L_3$ and V $L_2$ spectral intensity resonates at the main peaks at -11 eV and -2 eV, respectively. In V $L_{2,3}$ spectra of pure V (not shown), the peaks at -11 and -19 eV are absent and the main $L_3$ peak is located at -9 eV. The main peaks of the V $L_{2,3}$ SXE spectra of $V_2GeC$ and VC are quite different in shape. The difference is the relative intensity of the -11 eV substructure which is significantly higher in VC than in $V_2GeC$. This can be attributed to the pure octahedral coordination in VC in combination with very strong contribution of C $2p$ orbitals hybridizing into the V $3d$ valence band at -11 eV. The weak and broad structure observed around -19 eV for both $V_2GeC$ and VC (but not for pure V) is due to V $3d$ - C $2s$ hybridization at the bottom of the valence band. Note that for the resonant excitations at 516.9 eV and 522.1 eV, where the $2p$ core hole lifetime broadening (0.4 eV) has no influence, the V $L_{2,3}$ spectra of $V_2GeC$ excited with **E|a,b** polarization has a more similar shape to that of VC than the spectra excited with **E|c** polarization. On the contrary, the VC spectra do not

exhibit much polarization-dependent anisotropy as the V atoms in VC are isotropic with octahedral coordination symmetry. For the V$_2$GeC spectra, the anisotropy is attributed to the difference in coordination between the V atoms in-plane (octahedral coordination) and out-of-plane (trigonal prismatic coordination).

The V $L_{2,3}$ SXE spectra appear delocalized (wide bands) which makes electronic structure calculations particularly suitable for interpretation of nonresonant spectra. The calculated spectra (dotted curves at the bottom of Fig. 3) consist of V 3$d$ and 4$s$ pDOS obtained from full-potential *ab initio* density-functional theory projected by the 3$d$,4$s$→2$p$ dipole transition matrix elements and broadening corresponding to the experiment. The core-hole life-time broadening are set to 0.4 eV both for the 2$p_{3/2}$ and 2$p_{1/2}$ thresholds. To account for the $L_2$→$L_3M$ Coster-Kronig decay which change the initial core hole population from the statistical single-particle result (2:1), preceding the SXE process [17], the calculated spectra are fitted to the experimental $L_3/L_2$ ratio of 4.3:1. Furthermore, the 2$p_{3/2,1/2}$ spin-orbit splitting was set to the experimental SXE value of 8.2 eV. The 2$p_{3/2,1/2}$ core-level XPS spin-orbit splitting is 7.6 eV while the *ab initio* value is smaller (7.1 eV). The calculated *ab initio* values of the spin-orbit splittings in band-structure calculations are generally underestimated for the early transition metals (in this case 7.1 eV for V 2$p$) and overestimated for the late transition metals. The reason for this is not presently known, but must represent effects beyond effective one-electron theory e.g., many-body effects. In Fig. 3, the fitted spin-orbit splitting was set to the experimental $L_{2,3}$ SXE value of 8.2 eV while the 2$p_{3/2,1/2}$ core-level XPS spin-orbit splitting is 7.6 eV The integrated areas of the experimental and calculated spectra were normalized to the calculated V 3$d$+4$s$ charge occupations of V$_2$GeC: (3$d$: 2.452e, 4$s$: 2.067e), and VC: (3$d$: 2.389e, 4$s$: 2.070e). The area for the $L_2$ component was scaled down by the branching ratio and added to the $L_3$ component.

The final fitted spectra of V$_2$GeC, and VC are in fairly good agreement with the experimental results although the intensity distribution is somewhat different in the experiment. For VC, the $L_3$ double peak structure is in good agreement but a calculated peak at -4 eV is not present in the experiment which may partly be explained by the larger experimental 2$p_{1/2}$ lifetime broadening. For V$_2$GeC, the intensities of the -9 eV and -1 eV peaks are overestimated and the -11 peak underestimated in the calculation. Since the calculations do not include polarization effects, some of the differences in the intensities between experiment and theory may be attributed to the involvement of non-spherically symmetric V 2$p$ core states and 3$d$ valence states. However, part of the difference may also be due to the existence of the coherent VC$_x$ inclusions. The origin of the -11 eV peak is due to the $L_3$ component of a V 3$d$ pDOS peak at -3.3 eV below $E_F$ which is shifted -8.2 eV by the 2$p$ spin-orbit splitting. The much weaker 3$d$ pDOS contribution at -3.3 eV stemming from the $L_2$ component overlaps with the stronger $L_3$ emission line at -2 eV. The origin of the -9 eV peak is related to the $L_3$ component of a series of flat bands of V 3$d$ character resulting in high pDOS close to the $E_F$, shifted -8.2 eV by the 2$p$ spin-orbit splitting. The -2 eV peak, which is most intense at 522.1 eV excitation energy, corresponds to the $L_2$ component of the V 3$d$ pDOS, and the spectral shape is broader and less pronounced than the $L_3$ component due to the larger 2$p_{1/2}$ core-hole lifetime broadening.

## 4.2 C *K* x-ray emission

Figure 4 (top, right curves) shows C 1$s$ SXA spectra of V$_2$GeC and VC measured to identify the specific photon energy of the absorption intensity peak maxima. The SXA measurements mainly probe the unoccupied 2$p_{xy}$ in-plane orbitals. In the middle of Fig. 4, C *K* SXE spectra of V$_2$GeC and VC are shown. The **E**|**a,b** excited SXE spectra predominantly probe the occupied C 2$p_{x,y}$ in-plane orbitals. With **E**|**c** excitation, there is more influence from the out of plane C 2$p_z$ orbitals.

The main C K peak is located -3.8 eV below $E_F$ in both V$_2$GeC and VC. Comparing the C K SXE spectra to isostructural Ti$_2$AlC, the main peak is at -2.9 eV [9] a shift of +0.9 eV compared to the C K emission of V$_2$GeC (-3.8 eV). This is attributed to the additional valence charge density and core attraction in V compared to Ti. The higher C content in VC in comparison to V$_2$GeC, makes the C K spectra of VC somewhat broader than for V$_2$GeC with a steeper slope between 0 and -2 eV below $E_F$. The VC spectra indicate what the C $2p$ electronic structure of V$_2$GeC would look like if all Ge atoms would be exchanged by C atoms. The most important difference is the -6 eV shoulder which does not exist in VC. Although the C-Ge interaction is expected to be relatively weak, the additional shoulder at -6 eV in V$_2$GeC is attributed to C $2p$ - Ge $4p$ hybridization and charge-transfer mediated via the V $3d$ orbitals. The polarization dependence of the C K spectra is small and negligible for VC. However, a polarization dependence of the shoulder at -6 eV below $E_F$ in the resonant C K spectra of V$_2$GeC is observed. The -6 eV shoulder suggests an anisotropic C $2p$ hybridization with the Ge $4p$ orbitals as discussed in the next section.

For comparison of the peak intensities and energy positions, the integrated areas of the experimental and calculated spectra of V$_2$GeC and VC were normalized to the calculated C $2p$ charge occupations of V$_2$GeC:($2p$:

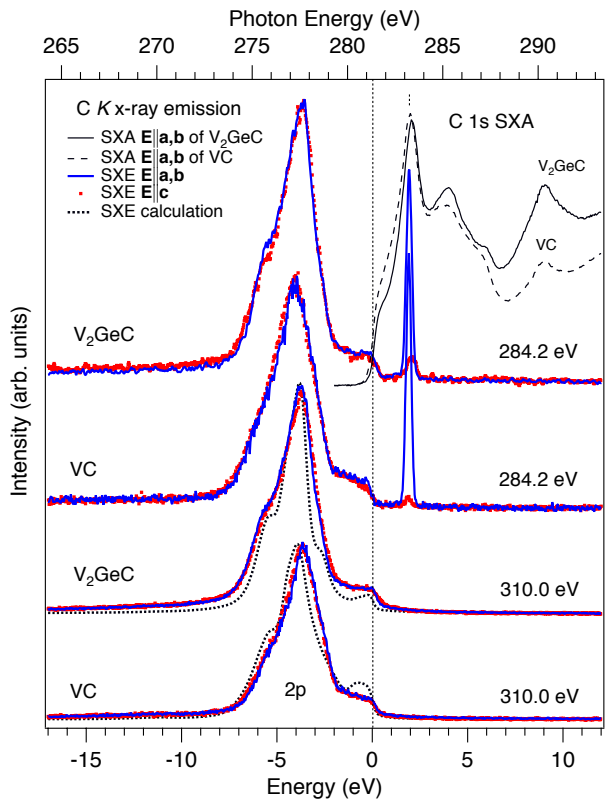

**Figure 4:** The top-right curves show C $1s$ SXA spectra measured at normal incidence (**E**|**a,b**), normalized to each other below and far above the absorption step edge. In the middle (top to bottom), C K SXE spectra of V$_2$GeC and VC are shown. The SXE spectra are excited with the polarization vector both parallel (**E**|**a,b**) and perpendicular (**E**|**c**) to the surface plane at 284.2 eV (resonant) and 310 eV (nonresonant). For each excitation energy, the spectra are normalized to the time and incoming photon flux from a gold mesh in the photon beam. The excitation energy of the resonant SXE spectra at **E**|**a,b** polarization (284.2 eV) is indicated by the strong elastic peaks determining the total resolution (FWHM=0.35 eV). The spectra are plotted both on a photon energy scale (top) and a common energy scale with respect to the $E_F$ (bottom) using the measured C $1s$ XPS binding energy of 281.2 eV. For the nonresonant spectra at the bottom, calculated isotropic SXE spectra are shown by the dotted curves.

2.360e) and VC:($2p$: 2.317e). Calculated C $K$ emission spectra including the C $2p$ pDOS projected by the $2p \rightarrow 1s$ dipole transition matrix elements and broadening corresponding to the experiment are shown at the bottom of Fig. 4. The general agreement between the experimental and theoretical spectra is good with a few exceptions. The predicted -6 eV shoulder in the VC spectra does not exist experimentally, neither at nonresonant not at resonant excitation energy. Moreover, the predicted peak structure at -0.5 eV in the calculated spectrum of VC is not visible experimentally.

## 4.3 Ge $M$ X-ray emission spectra

The top panel of Fig. 5 shows, experimental Ge $M_1$ SXE spectra, following the $4p \rightarrow 3s$ dipole transitions of V$_2$GeC and single crystal Ge (001). The probability for the Ge $M_1$ transitions is about two orders of magnitude lower than the Ge $M_{2,3}$ emission which makes the measurements demanding. For Ge in V$_2$GeC and pure Ge, the valence band is dominated by the Ge $4p_z$ and $4p_{xy}$ orbitals, oriented parallel and perpendicular to the surface normal **c** (see Fig. 1). Starting with V$_2$GeC, the $M_1$ spectra (within 20 eV from the E$_F$) exhibit two main peaks observed at -6 eV and -11 eV. For comparison, the $M_1$ spectra of pure Ge shown below, exhibit a peak at -2 eV with a shoulder at -6 eV. With **E**|**a,b** polarization, mainly the occupied Ge $4p_{xy}$ orbitals are probed. With **E**|**c** polarization, there is more influence from the out-of-plane $4p_z$ orbitals. The angular dependent anisotropy of the Ge $M_1$ emission is large at -6 eV in V$_2$GeC while it is absent in pure Ge. The intensity of the -6 eV peak is 3.7 times higher for **E**|**a,b** polarization than for **E**|**c** polarization. The absence of the -2 eV peak in V$_2$GeC may be caused by a combination of hybridization at lower energy, charge-transfer and final state screening. The peak structures observed at -32 to -34 eV in the $M_1$ spectra correspond to $3d \rightarrow 3s$ quadrupole transitions. These transitions are angular dependent with opposite peak intensities in comparison to the $3d \rightarrow 3p$ dipole transitions which will be discussed in connection to the lower panel. Due to the low transition probability, the quadrupole transitions exhibit a lower signal-to-background ratio than the dipole transitions. They appear at a somewhat lower energy than the corresponding dipole transitions in the lower panel which can be attributed to final state effects and screening. For comparison of the $M_1$ peak intensities and energy positions, the integrated areas of the experimental and calculated spectra of V$_2$GeC and VC were normalized to calculated Ge $4s$, $4p$, $3d$, and $4d$ charge occupations in V$_2$GeC:($4p$: 0.655e) and pure Ge:($4p$: 0.675e). The area for the $M_2$ component was scaled down by the experimental branching ratio and added to the $M_3$ component. The calculated Ge $M_1$ spectrum of V$_2$GeC was shifted by a -3.47 eV corresponding to the calculated screening effects by the energy difference between the $3d^9$ and $3d^{10}$ configurations.

The bottom panel of Fig. 5 shows experimental Ge $M_{2,3}$ SXE spectra of V$_2$GeC and Ge (001). The $4s$ valence band within 20 eV from the E$_F$ in the lower panel, consists of a weak double structure with $M_3$ and $M_2$ emission structures observed around -11 eV and -5 eV, respectively. As observed, the $4s$ valence-band structure is significantly more intense in V$_2$GeC than for pure Ge. Due to screening and charge-transfer, there is a -1.1 eV low-energy chemical shift of the Ge $3p_{3/2}$ and $3p_{1/2}$ XPS binding energies in V$_2$GeC (120.6 and 124.7 eV, respectively) in comparison to pure Ge (001) (121.7 and 125.8 eV, respectively). The -1.1 eV shift is significantly larger than the corresponding -0.3 eV shift observed for the Ge $3p_{3/2}$ and $3p_{1/2}$ XPS binding energies in Ti$_3$GeC$_2$ [20]. The shallow $3d$ core level has XPS binding energies of 28.4 eV and 29.6 eV for $V_2GeC$ and Ge, respectively. The $3d$ level consists of two peaks from the Ge $M_{4,5} \rightarrow M_3$ and $M_4 \rightarrow M_2$ ($3d \rightarrow 3p_{3/2,1/2}$) transitions with energies of -33.0 and -29.3 eV relative to the $E_F$. The measured $M_3/M_2$ intensity ratios of Ge in $V_2GeC$ are 1.65:1 with in-plane excitation and 1.92:1 with out-of-plane excitation. For pure Ge, the ratios are 1.65:1 for the in-plane excitation and 1.62:1 for the out-of-plane excitation which are all smaller than the statistical single-particle ratio of 2:1.

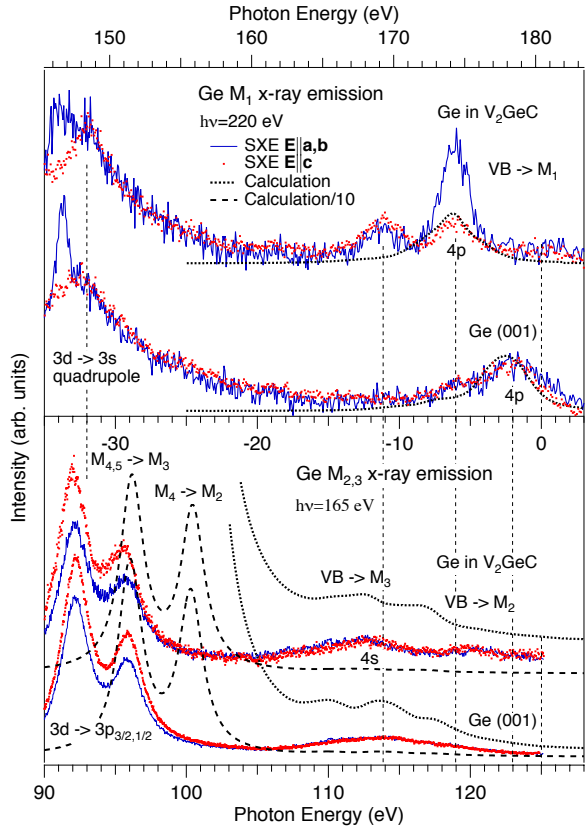

**Figure 5:** Experimental and calculated Ge SXE spectra of $V_2GeC$ and single-crystal Ge (001). Top panel: $M_1$ spectra nonresonantly excited at 220 eV (40 eV above the $3s$ threshold at 180.4 eV XPS binding energy) with the polarization vector parallel (**E**|**a,b**) and perpendicular (**E**|**c**) to the surface plane. A common energy scale relative to the $E_F$ is shown in the middle. Bottom panel: $M_{2,3}$ spectra nonresonantly excited at 165 eV (42 eV above the $3p_{1/2}$ threshold) with the polarization vector parallel (**E**|**a,b**) and perpendicular (**E**|**c**) to the surface plane. The spectra are plotted on a photon energy scale relative to the $E_F$ using the Ge $3p_{1/2}$ core-level XPS binding energy of 124.7 eV for $V_2GeC$ and 125.8 eV for Ge (001). Calculated isotropic SXE spectra are shown by the dotted curves. The calculated intensities for the shallow $3d$ core-levels were scaled down by a factor of 10 (dashed curves).

Calculated isotropic spectra are shown in the bottom panel in Fig. 5 by the full and dashed lines, both for the $4s$ valence band (full curves) and the $3d$ core levels (dashed curves). For comparison of the peak intensities and energy positions, the integrated areas of the experimental and calculated spectra of $V_2GeC$ and VC were normalized to the calculated Ge $4s$, $3d$, and $4d$ charge occupations in $V_2GeC$:($4s$: 0.764e, $3d+4d$: 9.867e,) and pure Ge:($4s$: 0.747e, $3d+4d$: 9.859e). For pure Ge, the calculated $4s$ DOS of the $M_3$ and $M_2$ valence bands are not only separated by the spin-orbit splitting of 3.6 eV, but are also split up into two subbands, separated by 3.5 eV (as in the case of single crystal bulk Si). The result is a triple structure in the valence band of Ge (001) where the upper and lower emission peaks are solely due to $M_3$ and $M_2$ emission bands, respectively, while the main middle peak is a superposition of both $M_3$ and $M_2$ contributions. On the contrary, the Ge $4s$ DOS in $V_2GeC$ consists of a single peak structure with more spectral weight towards the $E_F$ resulting in a double structure in the calculated $M_{2,3}$ emission. The measured Ge $M_{2,3}$ spin-orbit splitting of 3.6±0.1 eV [26] is smaller than the calculated value of 4.3 eV and the calculated 3*d* core levels (dashed curves) are also closer to the $E_F$ by 3.9 eV.

## 4.4 Chemical Bonding

Figure 6 shows calculated BCOOP [27] of $V_2GeC$ and VC. The calculated orbital overlaps makes it possible to compare the strength between covalent chemical bonds. A positive function below $E_F$ means covalent bonding states and a negative function above $E_F$ means anti-bonding states. The relative strength of the covalent bonding at a specific energy relative to the $E_F$ is determined by comparing the integrated areas under the curves. Increasing the energy distance from the $E_F$ of a positive bonding peak position also means that a larger strength of covalent bonding is achieved. Firstly, we note that the orbital overlaps of the ternary $V_2GeC$ are significantly more complicated than for the binary VC. For both systems, the main V 3*d* - C 2*p* overlap is found at -4 eV with additional peaks around -6 eV and -3 eV for $V_2GeC$. The V 3*d* - C 2*s* overlap has much lower intensity than the V 3*d* - C 2*p* overlap with non-covalent interaction at -11 eV and -3 eV. The V 3*d* - Ge 4*p* overlap has a large peak around -3 eV with additional smaller peaks at -3.5 eV, -4.5 eV and -6 eV in $V_2GeC$. We also note that the V 3*d* - Ge 4*p* overlap has filled bonding orbitals up to the $E_F$ while for the V 3*d* - C 2*p* overlap, also anti-bonding orbitals begin to be filled.

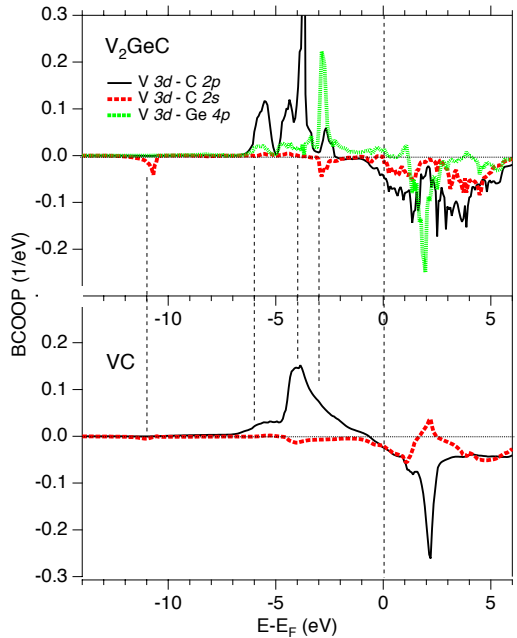

**Figure 6:** (Color online) Calculated balanced crystal overlap population (BCOOP) of $V_2GeC$ (top) and VC (bottom).

For the V $L_{2,3}$ SXE spectra of $V_2GeC$ and VC, discussed in section IV A, the BCOOP calculations show that strong V 3*d* - C 2*p* hybridization and covalent bonding takes place around -11 to -14 eV where charge is removed (-3 to -6 eV below $E_F$ when the 8.2 eV spin-obit splitting is subtracted). The V 3*d* - C 2*s* overlap is the origin of the -19 eV peak, the V 3*d* - Ge 4*p* overlap significantly contributes to the -11 eV peak and the -9 eV peak is mainly due to V 3*d* states. For the C *K* SXE spectra of $V_2GeC$ and VC, the main -3.8 eV peak and the -6 eV shoulder are due to strong V 3*d* - C 2*p* overlap and charge-transfer to C. For the Ge $M_1$ SXE spectra, V 3*d* Đ Ge 4*p* hybridization is predicted around -2.5 to -3 eV. However, contrary to the case of pure Ge, no peak structure is observed in this energy region in the Ge $M_1$ SXE spectra of $V_2GeC$. This is attributed to charge-transfer from the upper part of the valence band of Ge to the V 3*d* and C 2*s* orbitals. The enhanced -6 eV $M_1$ SXE peak in $V_2GeC$ can be interpreted as Ge 4*p* - V 3*d* orbital overlap with additional hybridization and charge-transfer from the C 2*p* orbitals. This interpretation is consistent with the additional -6 eV shoulder in the C 2*p* SXE spectra of $V_2GeC$ in comparison to VC. The -11 eV peak in the $M_I$ SXE spectra of V$_2$GeC appears at the same energy position as the -19 eV peak in the V 3$d$ spectra when the 8.2 eV V 2$p$ spin-orbit splitting is subtracted. However, the orbital overlap calculations show no sign of V 3$d$ - Ge 4$p$ interactions at -11 eV. The -11 eV peak structure probably originates from the non-covalent anti-bonding V 3$d$ - C 2$s$ interaction giving rise to charge transfer from the C 2$s$ orbitals to the Ge 4$p$ orbitals at the bottom of the valence band.

We find that the integrated intensity of the V 3$d$ - Ge 4$p$ BCOOP curve at the -2.9 eV peak is 16% larger than the corresponding Ti-Al peak at -0.64 eV in Ti$_2$AlC [9]. This shows that the V-Ge bonding in V$_2$GeC is stronger than the Ti-Al bonding in Ti$_2$AlC as also indicated by the shorter bond length in Table I. Note that the V-C bond length in V$_2$GeC is shorter than in the monocarbide VC. This is an interesting finding which has been observed in other MAX-phases [9, 10] and plays a key role for the physical properties of the material. However, due to the difference in atomic radii between V and Ti, there is no clear relationship between bond lengths and bond strengths in this case. Therefore it is more useful to compare V$_2$GeC with VC.

**Table 1:** Calculated bond lengths [Å] for V$_2$GeC, Ti$_2$AlC, VC and TiC. $M_I$ (octahedral coordination) is bonded only to C while $M_{II}$ (octahedral in-plane and trigonal prismatic coordination out-of-plane) is bonded to both C and the A-element as illustrated in Fig. 1.

| Bond type | $M_I$ - C | $M_{II}$ - C | $M_{II}$ - A | A-C |
|---|---|---|---|---|
| V$_2$GeC | - | 2.040 | 2.634 | 2.634 |
| Ti$_2$AlC | - | 2.117 | 2.901 | 3.875 |
| VC | 2.082 | - | - | - |
| TiC | 2.164 | - | - | - |

# 5 Discussion

The intercalation of Ge monolayers into the VC matrix mainly changes the character and scheme of the chemical bond regions as mentioned above, but also close to the $E_F$ where the conductivity occurs. The additional charge density introduced by hybridization with the Ge $sp$ bands leads to increased metallicity (decreased resistivity). The conductivity is largely governed by the V metal bonding and is roughly proportional to the total number of states at the $E_F$ (VC: 0.65 states/eV/atom, V$_2$GeC: 0.56 states/eV/atom, TiC: 0.12 states/eV/atom, and Ti$_2$AlC: 0.34 states/eV/atom). In reality, the V$_2$GeC film has significantly higher conductivity i.e., lower resistivity (0.21 μΩm) compared to Ti$_2$AlC (0.44 μΩm [28]). The resistivity of VC is even higher (0.93 μΩm) and for TiC an order of magnitude higher (2.00 μΩm [29]). From previous studies of ternary carbides [9], it is clear that those Ti atoms which are bonded to both C and the A element contribute more to the conductivity than the Ti atoms which bond solely to C. Since V$_2$GeC contains only V atoms bonded to both C and Ge, one would expect that V$_2$GeC has higher conductivity than ternary carbides with 312 and 413 crystal structures. In addition, there is a strong dependence of the valence electron concentration for the M-element.

In Figs. 3-6, we identified three main types of covalent chemical bonds, the strong V 3$d$ - C 2$p$ bond, the V 3$d$ - Ge 4$p$ bond, and the V 3$d$ - C 2$s$ bond. The V 3$d$ - C 2$p$ and V 3$d$ - C 2$s$ hybridizations are deeper in energy from the $E_F$ than the V 3$d$ - Ge 4$p$ hybridization indicating stronger bonding. Strengthening the M - A bonding should increase the stiffness of the material. A change of the elastic properties is achieved by the additional valence charge occupancy in V, in combination with the larger electronegativity and smaller atomic radius compared to Ti. A V 3$d$ - Ge 4$p$ chemical bond strengthening is indeed observed in the electronic structure of V$_2$GeC in comparison to Ti$_2$AlC. This is consistent with the calculated Youngs modulus ($E$-modulus) of V$_2$GeC (334 GPa) which is higher than for Ti$_2$AlC (305 GPa). However, the measured $E$-modulus is lower in V$_2$GeC (189 GPa) than in Ti$_2$AlC (260 GPa [30]). The difference between the experimental and calculated $E$-moduli can be due to the coherent VC$_x$ inclusions in the V$_2$GeC film. The measured $E$-moduli of both V$_2$GeC and Ti$_2$AlC are both significantly lower than for VC (255 GPa [31]) and TiC$_{0.8}$ (388 GPa [32]). Recently, it has been theoretically predicted that there is a linear relationship between the bulk moduli ($B$) of the ternary MAX-phases and the corresponding binary compounds as the valence band of the constituent M-element is being filled [34, 35]. Experimentally, for the $E$-modulus of single-crystal thin film materials, the corresponding linear relationship is more complicated to assess as measurements are made in a certain direction of the material. Moreover, for the binary transition metal carbides, it is theoretically known that the bond strength, bond energy and enthalpy decreases as the 3$d$ band of the constituent transition metal is being filled [36, 37].

# 6 Conclusions

In summary, we have investigated the anisotropy of the electronic structure of nanolaminate V$_2$GeC with the combination of polarization-dependent soft x-ray emission spectroscopy and electronic structure calculations. The measured emission spectra of V $L_{2,3}$, C $K$, Ge $M_1$ and Ge $M_{2,3}$ of V$_2$GeC were compared to monocarbide VC and pure Ge. Ge $M_1$ x-ray emission spectra of V$_2$GeC and pure Ge were presented. Different types of covalent chemical bond regions were revealed; V 3$d$ - C 2$p$ bonding at -3.8 eV, Ge 4$p$ - C 2$p$ bonding at -6 eV and Ge 4$p$ - C 2$s$ interaction mediated via the V 3$d$ orbitals at -11 eV below the Fermi level. We found that the anisotropic effects were most pronounced for the Ge 4$p$ valence states and the shallow Ge 3$d$ core levels, while relatively small anisotropy was detected for the V 3$d$ states. In comparison to single-crystal Ge, the $M_1$ x-ray emission spectra of V$_2$GeC exibit more intensity and charge-transfer to the $4p_{xy}$ in-plane valence-band orbitals than to the $4p_z$ out-of-plane orbitals. The anisotropic pattern of the chemical bond regions as revealed for V$_2$GeC determines the electrical and thermal conductivity and the elastic properties of the material. As shown, measuring the polarization-dependence of the electronic structure can be utilized as an advanced tool to probe the internal orbital hybridization with different symmetries and rigid bond directions in nanostructured materials.

# 7 Acknowledgements

We would like to thank the staff at MAX-lab for experimental support. This work was supported by the Swedish Research Council, the Göran Gustafsson Foundation, the Swedish Strategic Research Foundation (SSF), Strategic Materials Reseach Center on Materials Science for Nanoscale Surface Engineering (MS$^2$E), and the Swedish Agency for Innvovations Systems (VINNOVA) Excellence Center on Functional Nanostructured Materials (FunMat). One of the authors (M. Mattesini) wishes to acknowledge the Spanish Ministry of Science and Technology (MCyT) for financial support through the *Ramón y Cajal* program.